\documentclass[cameraready]{acmsiggraph}

\title{The Graphics Card as a Stream Computer}
\author{Suresh Venkatasubramanian \\ AT\&T Labs -- Research} 
\begin{document}
\maketitle

\newcommand{\note}[1]{{\Large\sc #1}}
\newcommand{\etal}{\emph{et al.}}
\section{Introduction}
\label{sec:intro}

Massive data sets have radically changed our understanding of how to
design efficient algorithms; the \emph{streaming paradigm}, whether it in
terms of number of passes of an external memory algorithm, or the single
pass and limited memory of a stream algorithm, appears to be the dominant
method for coping with large data. 

A very different kind of massive computation has had the same effect at
the level of the CPU. It has long been observed~\cite{B78} that the
traditional Von Neumann-style architecture creates memory bottlenecks
between the CPU and main memory, and much of chip design in the past many
years has been focused on methods to alleviate this bottleneck, by way of
fast memory, caches, prefetching strategies and the like. However, all of
this has made the memory bottleneck itself the focus of chip optimization
efforts, and has reflected in the amount of real estate on a chip devoted
to caching and memory access circuitry, as compared to the ALU
itself~\cite{D03}.

For compute-intensive operations, this is an unacceptable tradeoff. The
most prominent example is that of the computations performed by a graphics
card. The operations themselves are very simple, and require very little
memory, but require the ability to perform many computations extremely
fast and in parallel to whatever degree possible. Inspired in part by
dataflow architectures and systolic arrays, the development of
graphics chips focused on high computation throughput while sacrificing
(to a degree) the generality of a CPU. 

What has resulted is a \emph{stream processor} that is highly optimized
for stream computations. Today's GPUs (graphics processing units) can process over
50 million triangles and 4 billion pixels in one second. Their ``Moore's Law''
is faster than that for CPUs, owing primarily to their stream architecture
which enables all additional transistors to be devoted to increasing
computational power directly. 
An intriguing side effect of this is the growing use of a graphics card as
a general purpose stream processing engine. In an ever-increasing array of
applications, researchers are discovering that performing a
computation on a graphics card is far faster than performing it on a CPU,
and so are using a GPU as a stream co-processor. 
Another feature that makes the graphics pipeline attractive (and
distinguishes it from other stream architectures)
is the \emph{spatial parallelism} it
provides. Conceptually, each pixel on the screen can be viewed as a stream
processor, potentially giving a large degree of parallelism.

\section{A Model Of A Graphics Card}
\label{sec:graphics}

A detailed description of the architecture of a graphics card is beyond
the scope of this note; what follows is a distillation of the key
components. 
The stream elements that a card processes are points, and at a later
stage, \emph{fragments}, which are essentially pixels with color and depth
information. A program issues requests to the card by specifying points,
lines, or facets, (and their attributes) and issuing a rendering
request. The vertices go thru several processing phases (called a
\emph{vertex program}) and finally are
scan-converted into fragments by a \emph{rasterizer} that takes a
three-dimensional scene and breaks it into pixels. 
Each fragment is then processed by another stream computation (called a
\emph{fragment program}) and finally ends up as a pixel on a
screen. Alternatively, the pixels and their attributes can be captured in
internal storage and their contents can be retrieved to the CPU. 
Spatial parallelism occurs at the level of a fragment; each fragment
program can be viewed as running in parallel at each pixel. The vertex
programs are executed on each point. \emph{Texture memory} provides a form
of limited local storage for maintaining state; access to it is restricted
though.

\subsection{A Computational Model}

From the above description, it is clear that a GPU executes a stream
program. There are some key differences between it and a general purpose
stream program. The most essential one is that
\emph{every object in the stream is processed by the same function}

Recall that in a general stream computation~\cite{FKSV99} the data streams by an
\emph{arbitrary} Turing machine that has access to limited memory. There
is no restriction in general on what is computed; the restriction that the
hardware enforces is thus significant. Further, the computations
themselves are limited to branching programs; no looping or
jumping is allowed\footnote{there are some exceptions, but these are not
  significant}. 

Another important (and related) difference is the access to memory; a
stream program on a graphics card has access to much smaller memory than a
general stream program. Although graphics cards have a great deal of
memory on them (256 MB in the best cards available today), each processing
unit really has access to only a few fast registers, unlike the much
larger memory limits on a standard streaming process. 

Finally, the typical computation performed in the GPU is multi-pass; a
constant or even $\omega(1)$ passes may be required for a particular
computation.

\subsection{Less Is More}
\label{ssec:less}

Although it seems that the above restrictions are artificial and limiting,
it turns out they are at the heart of the performance gains achieved by
graphics cards. Because each item in the stream is processed in the same
way, microcode for the computation can be loaded directly onto the GPU,
and no expensive memory access are needed to fetch the next instruction.
Using branching programs enforces the pipeline discipline, and ensures
that items are processed without stalls. Lack of memory access allows
small caches to be used and prevents the memory bottleneck from swamping
the chip.  It is also important to observe that other streaming
architectures that have been designed and proposed (\cite{I,C}) share little
 in common with graphics cards but share the above
\emph{uniformity} assumption. Thus, uniform streaming appears to be the
common underlying computational metaphor for stream architectures in general.

\paragraph{Limitations On Computing Power}
As far back as 1989, Fournier and Fussell~\shortcite{FF88} presented a view of
the GPU as a stream computing engine, and presented various upper and
lower bounds for simple problems. They show for example that sorting is a
hard problem in this model, and investigate how the power of the pipeline
changes as more and more registers and operations are added to the model.

More recently, Guha~\etal~\shortcite{GKMV03b} showed that computing the median
of $n$ numbers takes $n$ passes in the ``standard'' graphics pipeline
model. This contrasts to $O(\log n)$ passes required to compute the median
in a stream model. Interestingly, the addition of an extra
operator (an interval test of the form ``Given z, is $a \le z \le b$'')
reduces the number of passes needed (randomized) to $O(\log n)$.

\section{Examples}
\label{sec:pipeline}
There are numerous examples of places where the GPU has been used to
perform general stream computations. A somewhat incomplete list is below:
\begin{itemize}
\item General distance field calculations: Given a distance function
  and a set of objects, compute for each object its
  (nearest/farthest/median) neighbour. Yields algorithms for Voronoi
  diagrams~\cite{HKLMC99}, general shape fitting~\cite{AKMV02}, spatial
  convolution and  outlier detection and other problems. 
\item Object intersections: given two objects, determine if they
  intersect and by how much. Yields algorithms for spatial joins~\cite{SAA03},
  penetration depth~\cite{AKMV02}, range searching~\cite{KMMV02}.
\item Scientific computing: Solve partial differential equation
 via finite element methods~\cite{DPRS01}
\item Physical simulation: Simulate the motion of objects in a physically
  realistic way. Useful for game programming and dynamical system modelling.
\end{itemize}

It is worth noting that none of the above applications make use of the
extensive features of vertex and fragment programs, and in fact only use a
very primitive subset of operations.

\section{Directions}
\label{sec:concl}

There are a number of challenging directions to pursue. There is already
work on software and language constructs to express general purpose stream
programs in terms of low-level graphics hardware commands (so that the
programmer need know nothing about graphics)~\cite{BH03}. There is work in
compilers on optimizing the translation of high-level stream constructs~\cite{C02}.
Other work seeks to extend the uniform streaming model to allow limited
caching of state~\cite{D03}.

From a computational standpoint, all of this leads to a fundamental
question: What is the power of uniform streaming and its variants ? The
answer will be invaluable not only from a theoretical point of view, but
as a way of indicating what problems are likely to be amenable to
efficient stream computation.

\bibliographystyle{acmsiggraph}
\bibliography{hardware}

\begin{thebibliography}{\protect\citename{Feigenbaum et~al\mbox{.} }1999}

\bibitem[\protect\citename{Agarwal et~al\mbox{.} }2002]{AKMV02}
{\sc Agarwal, P., Krishnan, S., Mustafa, N., and Venkatasubramanian, S.}
\newblock 2002.
\newblock Streaming geometric computations using graphics hardware.
\newblock Tech. rep., AT\&T Labs--Research.

\bibitem[\protect\citename{Backus }1977]{B78}
{\sc Backus, J.}, 1977.
\newblock Can programming be liberated from the von {N}eumann style ? a
  functional style and its algebra of programs.
\newblock ACM Turing Award Lecture.

\bibitem[\protect\citename{Bove and Watlington }1995]{C}
{\sc Bove, V., and Watlington, J.}, 1995.
\newblock Cheops: A reconfigurable data-flow system for video processing.

\bibitem[\protect\citename{Buck and Hanrahan }2003]{BH03}
{\sc Buck, I., and Hanrahan, P.}, 2003.
\newblock Data parallel computation on graphics hardware.
\newblock Manuscript.

\bibitem[\protect\citename{Chan et~al\mbox{.} }2002]{C02}
{\sc Chan, E., Ng, R., Sen, P., Proudfoot, K., and Hanrahan, P.}
\newblock 2002.
\newblock Efficient partitioning of fragment shaders for multipass rendering on
  programmable graphics hardware.
\newblock In {\em Proc. SIGGRAPH/Eurographics Workshop on Graphics Hardware}.

\bibitem[\protect\citename{Diewald et~al\mbox{.} }2001]{DPRS01}
{\sc Diewald, U., Preusser, T., Rumpf, M., and Strzodka., R.}
\newblock 2001.
\newblock Diffusion models and their accelerated solution in computer vision
  applications.
\newblock {\em Acta Mathematica Universitatis Comenianae (AMUC)\/}.

\bibitem[\protect\citename{Duca et~al\mbox{.} }2003]{D03}
{\sc Duca, N., Cohen, J., and Kirchner, P.}, 2003.
\newblock Stream caching: A mechanism to support multi-record computations
  within stream processing architectures.
\newblock DIMACS Working Group on Streaming Analysis II, March.

\bibitem[\protect\citename{Feigenbaum et~al\mbox{.} }1999]{FKSV99}
{\sc Feigenbaum, J., Kannan, S., Strauss, M., and Viswanathan., M.}
\newblock 1999.
\newblock An approximate l1-difference algorithm for massive data streams.
\newblock In {\em Proc. 40th Symposium on Foundations of Computer Science},
  IEEE.

\bibitem[\protect\citename{Fournier and Fussell }1988]{FF88}
{\sc Fournier, A., and Fussell, D.}
\newblock 1988.
\newblock On the power of the frame buffer.
\newblock {\em ACM Transactions on Graphics\/}, 103--128.

\bibitem[\protect\citename{Guha et~al\mbox{.} }2002]{GKMV03b}
{\sc Guha, S., Krishnan, S., Munagala, K., and Venkatasubramanian, S.}
\newblock 2002.
\newblock The power of a two-sided depth test and its application to csg
  rendering and depth extraction.
\newblock Tech. Rep. TD-5FDS6Q, AT\&T.

\bibitem[\protect\citename{{Hoff {III}} et~al\mbox{.} }1999]{HKLMC99}
{\sc {Hoff {III}}, K.~E., Keyser, J., Lin, M., Manocha, D., and Culver, T.}
\newblock 1999.
\newblock Fast computation of generalized {Voronoi} diagrams using graphics
  hardware.
\newblock {\em Computer Graphics 33}, {Annual Conference Series}, 277--286.

\bibitem[\protect\citename{Kapasi et~al\mbox{.} }2002]{I}
{\sc Kapasi, U.~J., Dally, W.~J., Rixner, S., Owens, J.~D., and Khailany, B.}
\newblock 2002.
\newblock The imagine stream processor.
\newblock In {\em Proc. IEEE International Conference on Computer Design},
  282--288.

\bibitem[\protect\citename{Krishnan et~al\mbox{.} }2002]{KMMV02}
{\sc Krishnan, S., Mustafa, N., Muthukrishnan, S., and Venkatasubramanian, S.}
\newblock 2002.
\newblock Extended intersection queries on a geometric {SIMD} machine model.
\newblock Tech. rep., AT\&T.

\bibitem[\protect\citename{Sun et~al\mbox{.} }2003]{SAA03}
{\sc Sun, C., Agrawal, D., and Abbadi, A.~E.}
\newblock 2003.
\newblock Hardware acceleration for spatial selections and joins.
\newblock In {\em SIGMOD}.

\end{thebibliography}

\end{document}